# Building Blocks of a User Experience Research Point of View

Building Blocks of a UXR POV

Real life examples

Patricia Diaz

Google, patdiaz@google.com

This paper presents three User Experience Research (UXR) perspectives based on data, evidence and insights –known as Point of View (POV)– showcasing how the strategies and methods of building a POV work in an enterprise setting. The POV are:

1. Smart Visuals: Use AI to extract and translate text from visuals in videos (2019).

2. Assessable Code Editor: Focus on direct AI-feedback to the learner as it is the loop that requires the least effort for the highest impact (2023).

3. Opportunity Landscape: Identify high-impact opportunities at the intersection of emergent technical capabilities that unlock novel approaches to critical user needs while addressing business strategic priorities (2019).

They all seemed far-fetched and went against common practice. All were adopted and had long-lasting impact.

## 1 INTRODUCTION

The UXR POV Playbook provides a set of criteria, standards, and tools to guide multidisciplinary teams in designing products and services [1]. A POV can go a long way by influencing influencers especially when putting forward a proof-of-concept of a novel idea –against the establishment– that has, otherwise, little chance to crystalize. For example, the multilingual interfaces of platforms that we now take for granted were not part of the User Interface (UI) design repertoire back in the early 2000s where "mirror" translated sites prevailed. When I led the development of the Clubhouse Network social network (referred to as "intranet" back then, and fondly named "the Village"), I thought it was imperative to make it multilingual in order to adhere to our guiding principles.

Having Clubhouses in several continents meant, among other things, embracing a myriad of languages and cultures. Even though the Network language is English, the Village strives to be a multilingual community where members are welcome to participate in any language they feel comfortable using. We facilitated this interchange by providing an interface in a dozen languages. This allows participants to see the same content, while navigating in their own language. Our POV was that a multilingual audience deserved a virtual meeting space where people were welcome to communicate using diverse languages, just like in the Village green of a multilingual town. This oriented design decisions such as choosing common icons familiar to multiple cultures and relying on them as much as possible in place of text. The interface was, of course, just the beginning. We fostered a multilingual culture, where people felt compelled to communicate beyond language barriers [2].

## 2 FOUNDATION

At the base of the pyramid, it is vital to establish clear business objectives and a robust research project purpose and plan. This forms the bedrock of POV credibility [1]. By the end of 2019, there were many research projects with promising AI capabilities: at Google we had advanced Optical Character Recognition (OCR), label detection, translation, text-to-speech, and speech-to-text. In a preview of their 2020 roadmap, HapYak executives made us aware of their new machine vision research for interactive videos. If we could put all those pieces together we would be able to attend our leadership call to "create screencasts for practitioner content that will be easier to localize".

Without knowing if it was even possible to combine them, let alone in the ways that we dreamed, there was no point in considering other building blocks of a POV (see Figure 1). However, the business objective was utterly clear: we had a commitment to start producing our educational content in at least 20 languages in addition to English without altering our delivery timeline and the only solution at hand implied a huge compromise that would affect our final users. The upfront investment in foundational research was worth it and we had a clear plan: only if the results of our proof-of-concept demonstrated feasibility we would move to data collection, insight generation, and ultimately a POV.

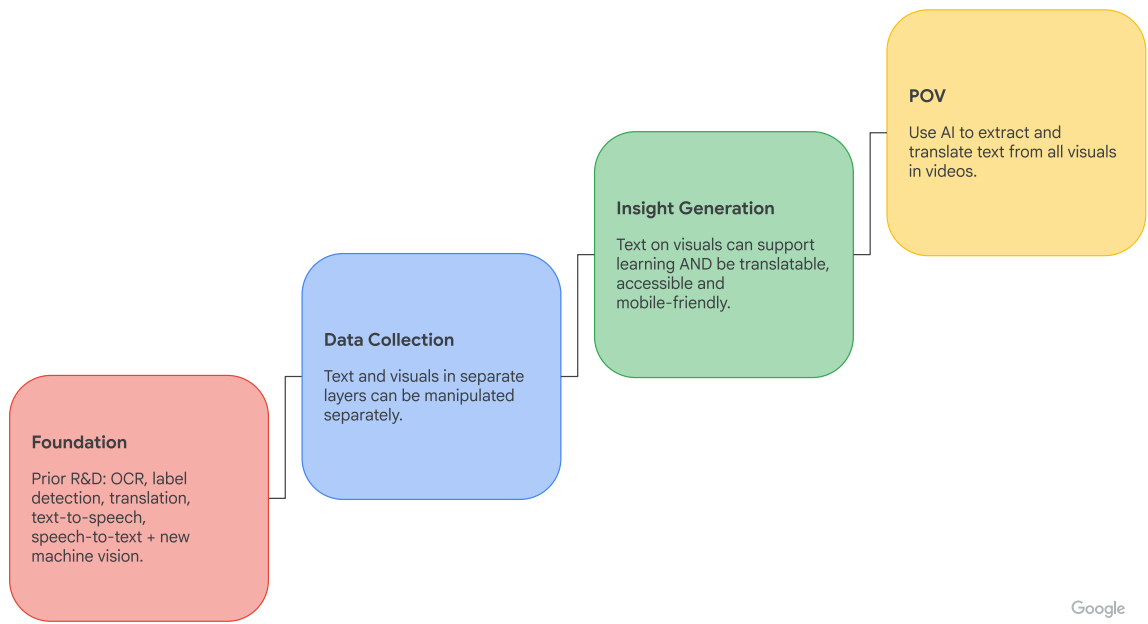

Figure 1: Smart Visuals POV building blocks

Because we had no easy way to translate the rich infographics we were producing to facilitate learning of crucial concepts, we had been asked to stop using text on top of graphics. In other words, we had to choose between making *concepts* more accessible or allowing our users to interact with our content in their preferred language and device while complying with traditional accessibility guidelines.

The prototype was ready in a few months and performed much better than expected but raised concerns of potential damage to our image and reputation given the unusual approach of overlaying the translation on the original text. Though this is now a common design pattern, we moved quickly to collect data to explore how it was perceived at the moment when we had little to no precedent. Our users were delighted and with the insights we were able to generate, our partner company moved into full production mode.

The resulting POV was straightforward: use AI to extract and translate all text on visuals. What started for Hapyak as an added feature became the core of a new product that helped them get acquired. It was ready at a time when localization became paramount for school systems as part of following new regulations. On May 2021, they were finalists for the Education Technology CODiE Awards in the category of "Best Use of Artificial Intelligence in EdTech" [3].

## 3 DATA COLLECTION

The process of creating a POV involves gathering pertinent data, both existing and new, utilizing suitable research resources and techniques. Data quality remains central, as it directly influences the generation of meaningful insights [1]. In 2022, we set out to evaluate the feasibility of collecting signals from software developers as they learned new aspects of coding in the Integrated Development Environment (IDE), and the usefulness of those signals to generate insights to inform the production of future learning materials.

There was plenty of foundational learning engineering research [4] that we complemented with our own small effort, so our main focus from the start was on data collection. I created a suite of studies to trace the signals collected in the developers logs which was very straightforward, and also devised a way to measure the usefulness of those signals without actually building the infrastructure to process them. With immense help from learning researchers and a data scientist, I ran a cognitive walkthrough before collecting data from developers –precisely to identify what data was valuable to collect–, and a second cognitive walkthrough to evaluate if the data collected was enough to make decisions about future content (see Figure 2).

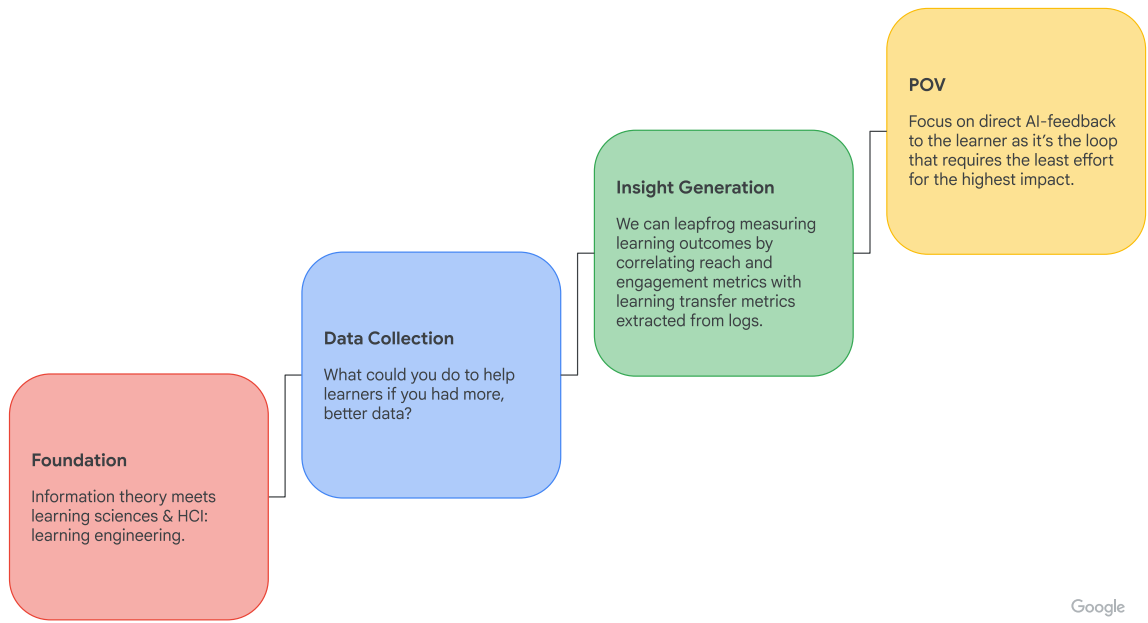

Figure 2: Assessable Code Editor Pilot POV building blocks

Along the way, the lead engineer, Davy Risso, opportunistically implemented Machine Learning (ML) feedback directly to the learner that turned out to be a game-changer. One of the insights was that we can leapfrog measuring learning outcomes by correlating reach and engagement metrics with learning transfer metrics extracted from the logs. It was also evident that the direct feedback to learner using Artificial Intelligence (AI) held a lot of promise, even if the Large Language Models (LLMs) were not fully ready at the time.

The very unexpected POV that we should invest in AI-powered assistance for learning jump-started rapid prototyping with AI in our team.

## 4 INSIGHT GENERATION

Moving upward, the focus is on uncovering impartial meanings within the data, which serves the dual purpose of informing business decisions and empowering users [1]. In 2019, our team stopped using a collage of disjointed tools from a myriad vendors and executed on a vision to create a unified technological platform that addressed the needs of designers and final users.

With our large team of highly-skilled professionals, many coming from academia, practice built on robust research – including Jobs-to-be-done [5, 6]– and data was collected on a daily basis as part of our workflow. Through an unorthodox way of looking at bug reports, I noticed that there was a big discrepancy given by what users expected and what developers considered was working as intended. It was the classic conundrum of users considering a bug what developers thought was a feature (see Figure 3).

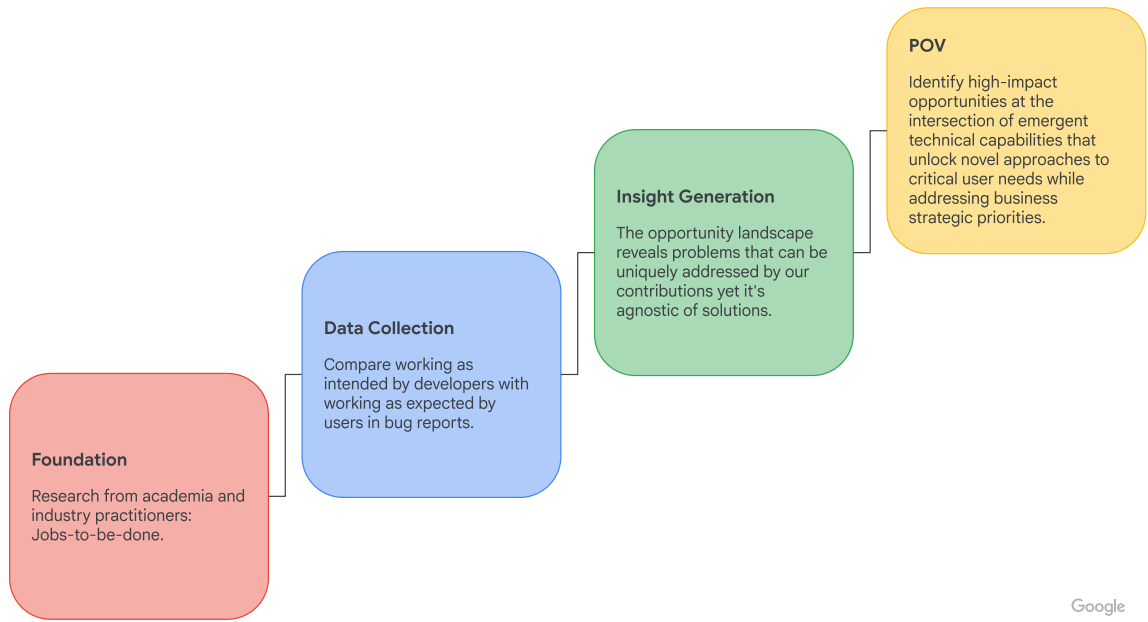

Figure 3: Opportunity Landscape POV building blocks

The insight required a lot of self-education –happily embraced by our team– to fully understand what the data was telling us and to incorporate its implications into our processes. The organic and rapid growth of our team had resulted in bespoke solutions that were addressing niche use cases in detriment of a holistic approach. As a result, some of our users who were the beneficiaries of more than one custom path, were baffled.

## 5 CONCLUSION

Innovation demands wide adoption which, in turn, requires that others identify with our ideas so deeply that at times they forget our intervention and present them as their own. More often than not, innovation is the result of many iterations; sometimes they become part of the global imaginary only after a final touch that builds on the work of many who created the bulk of the product. Those who give the final touch may be the only ones remembered, yet in many cases, what ignited the whole effort was a well-crafted POV.

### ACKNOWLEDGMENTS

Many people contributed to the efforts shared. It would be impossible to name all the individuals. I'm grateful for the countless learnings that came from the interactions with Clubhouse personnel and youth around the world, colleagues at the Museum of Science, Boston, MA; MIT Media Lab, Cambridge, MA; Hapyak, Boston, MA; and many Google offices.